**Isolated spin qubits in SiC with a high-fidelity infrared spin-to-photon interface**

David J. Christle[1], Paul V. Klimov[1], Charles F. de las Casas[1], Krisztián Szász[2], Viktor Ivády[2,3], Valdas Jokubavicius[3], Jawad ul Hassan[3], Mikael Syväjärvi[3], William F. Koehl[1], Takeshi Ohshima[4], Nguyen T. Son[3], Erik Janzén[3], Ádám Gali[2,5], David D. Awschalom[1]

1. Institute for Molecular Engineering, University of Chicago, Chicago, IL 60637, USA
2. Wigner Research Centre for Physics, Institute for Solid State Physics and Optics, Hungarian Academy of Sciences, PO Box 49, H-1525, Budapest, Hungary
3. Dept. of Physics, Chemistry and Biology, Linköping University, SE-581 83 Linköping, Sweden
4. National Institutes for Quantum and Radiological Science and Technology, 1233 Watanuki, Takasaki, Gunma 370-1292, Japan
5. Dept. of Atomic Physics, Budapest University of Technology and Economics, Budafoki út 8., H-1111, Budapest, Hungary

**Abstract**

The divacancies in SiC are a family of paramagnetic defects that show promise for quantum communication technologies due to their long-lived electron spin coherence and their optical addressability at near-telecom wavelengths. Nonetheless, a mechanism for high-fidelity spin-to-photon conversion, which is a crucial prerequisite for such technologies, has not yet been demonstrated. Here we demonstrate a high-fidelity spin-to-photon interface in isolated divacancies in epitaxial films of 3C-SiC and 4H-SiC. Our data show that divacancies in 4H-SiC have minimal undesirable spin-mixing, and that the optical linewidths in our current sample are already similar to those of recent remote entanglement demonstrations in other systems. Moreover, we find that 3C-SiC divacancies have millisecond Hahn-echo spin coherence time, which is among the longest measured in a naturally isotopic solid. The presence of defects with these properties in a commercial semiconductor that can be heteroepitaxially grown as a thin film on Si shows promise for future quantum networks based on SiC defects.

**Main Text**

**Introduction**

Deep-level defects in semiconductors are attractive systems for exploring quantum information and quantum-metrology applications. Similar to atoms, defects can possess electronic spin states with long coherence times that can be initialized and detected optically, but they also have the advantage of being fixed within a solid. The nitrogen-vacancy (NV) center in diamond is the most prominent deep-level defect for advanced quantum optical experiments, and has a rare combination of single-defect addressability, long spin coherence times, and a good optical interface. These properties have been established over five decades of research, and are at the foundation of cutting-edge demonstrations of long distance photon-mediated entanglement, teleportation, and the loophole-free test of Bell's inequalities [1–3].

In recent years, attention has increasingly been directed towards finding optically active defects with quantum mechanical properties that would complement or improve on those seen in NV centers. In particular, the NV's visible-wavelength emission and the difficultly of growing and microfabricating diamond are significant barriers to long-distance entanglement and creating scalable, integrated nanophotonic structures. Silicon vacancies and nickel centers in diamond [4,5], rare earth and transition metal ions [6–8], and several different vacancy-related complexes in silicon carbide [9–14], are all examples of possible alternatives that are being explored. These defects can have certain attractive features, but in many cases they have been controlled only in ensembles, have relatively short spin coherence times, or suffer from low initialization and readout fidelities. Even with compelling fundamental or materials advantages, it is difficult to assess a defect's potential role in future quantum technologies until the key milestones that overcome these challenges are reached.

In this work, we meet three of these milestones for divacancy defects in silicon carbide (SiC) – we isolate single defects in the cubic 3C polytype of SiC for the first time, demonstrate that their spin-coherence times are long, and reveal a robust spin-to-photon interface in both 3C- and 4H-SiC divacancies that can be used for high-fidelity preparation and readout using resonant infrared light. Each of these is a vital component for entangling remote defects, and establishing them makes a compelling case for utilizing the distinct advantages of SiC divacancies for long-distance entanglement and quantum networks. An exciting result is that its optical fine structure is akin to that of the diamond NV center, which means that the high-fidelity preparation and readout protocols used there will translate, so long as certain technical figures of merit are similar. In our experiments in 4H-SiC, the optical linewidths and quality of the cycling transitions available for spin readout are already near those used in state-of-the-art NV experiments.

Divacancy defects in SiC have long-lived electronic spin-triplet ground states that can be polarized and detected using light [12,15,16], similar to NV centers in diamond. However, the divacancies have several potential advantages, particularly in the context of photonics. Namely, they are optically active near the telecom wavelengths, and their host material, SiC, is available commercially as low-impurity, single-crystal wafers up to several inches in diameter, and is amenable to many existing device fabrication protocols [17,18]. In addition, 3C-SiC can be grown as an epitaxial film on Si, which provides a natural way to obtain thin membranes for photonics and micromechanics applications rather than the thinning of bulk crystals [19–22]. Incorporating good spin qubits into a photonic structures has been a longstanding area of interest [23,24], and this has partly driven the substantial amount of research in studying the spin and optical properties of SiC divacancy ensembles over the last few years [12,25–27]. Recently, single divacancies with millisecond spin coherence times were isolated in 4H-SiC, which allows the control of single quantum states and the emission of single photons [28]. If divacancies possess a highly spin-dependent spin-photon interface, these photons could be used to coherently link divacancy spins over long distances via optical fiber [1,29,30], but only low-fidelity conversion between spin and light has been demonstrated in SiC defects thus far [28,31]. Furthermore, achieving practical photon-mediated entanglement rates in defects, including the divacancy, will likely require integration with photonic structures that improve their spectral indistinguishability and photon emission rate. The general device-friendliness of SiC may ease the fabrication of these structures [32], but no single defect spins have been isolated specifically in 3C-SiC, where thin membranes are easiest to create.

**ISOLATION OF SINGLE DIVACANCIES IN 3C-SiC**

Divacancies in SiC are a point defect consisting of a missing Si atom adjacent to a missing C atom. SiC may exist in a variety of different polytypes, but the most commonly studied and readily available polytypes are labeled 4H-, 6H-, and 3C-SiC. These polytypes have different stacking periodicity and crystalline structure, and, as a result, a different number of inequivalent divacancies exist in each polytype. There is a single form of divacancy in 3C-SiC, four inequivalent forms in 4H-SiC, and six in 6H-SiC. Ensembles of each inequivalent divacancy form in its neutral charge state have been detected using optically detected magnetic resonance (ODMR) in 4H- and 6H-SiC using their distinct ground state spin resonances [33].

Although 3C has thus far received less attention for coherent spin control than 4H and 6H, it may be a very important polytype for photonics and opto-mechanical applications because it can be grown as an epitaxial thin film on Si [21,34–36]. An EPR defect labeled "Ky5" [37] and the ODMR center labeled "L3" [38] in 3C-SiC have similar magnetic and optical properties to the predicted 3C-SiC neutral divacancy [39] and those observed in the hexagonal polytypes [33], but only tentative assignment as the neutral divacancy has been made.

To isolate single Ky5/L3 defects, we start with a thick (~1.5 mm), single crystal and polytype inclusion-free bulk-like layer of 3C-SiC grown by sublimation on 4H-SiC substrate. This layer has significantly better epitaxial matching and lower density of structural defects compared to 3C-SiC material grown on Si substrates [40]. After removing the 4H-SiC substrate and the defective interface region by mechanical polishing, we obtain a 730 μm thick freestanding 3C-SiC layer. In order to isolate single Ky5/L3 defects, we irradiate the sample with 2.5 MeV electrons at room temperature to a fluence of 5 x $10^{12}$ $cm^{-2}$ and perform a 30 minute anneal in Ar gas at 745 °C (see Supplementary Information). Spatially-resolved PL scans using about 1 mW of off-resonant 975 nm laser excitation show isolated bright spots (Fig. 1a) that have a zero-phonon-line (ZPL) at 1106 nm (see Supplementary Information), which is consistent with previous Ky5 and L3 reports [33,38]. Photon coincidence correlation measurements (Fig. 1b) on several of these spots show an antibunching signal consistent with an isolated quantum emitter [41], which proves that they are from single defects.

We measure continuous-wave ODMR on these single defects by sweeping an applied microwave frequency and monitoring the PL emitted under off-resonant laser excitation. The spin resonances we detect are consistent with a spin-1 defect with a zero-field-splitting (ZFS) of $D$ = 1.336 GHz. This is in agreement with previous Ky5 reports and shows that the L3 defect that has been detected in ODMR and the Ky5 defect that has been detected in EPR are the same defect. To demonstrate coherent control of these single spins, we detect coherent Rabi oscillations by initializing and detecting the spin using ~1 μs laser pulses, and applying microwave bursts between pulses for spin rotation (Fig. 1c). These 7.5% Rabi readout contrast we measure is similar to the 9-15% contrasts seen in different forms of single 4H-SiC divacancies [28].

We characterize the inhomogeneous spin coherence times ($T_2^*$) of a few of these defects using Ramsey spectroscopy. The longest we observe is $T_2^*$ = (1.8 ± 0.1) μs, and the shortest is (250 ± 90) ns on a different defect (see Supplementary Information). Differences in the local impurity density may explain this variation. To probe their homogeneous coherence time, we apply a Hahn echo pulse sequence on a second sample that contains an ensemble of Ky5 defects created with a large 1 x 10$^{15}$ cm$^{-2}$ irradiation fluence. This measurement sets a lower bound on the single-spin homogeneous coherence, and the larger signal makes such a long-duty-cycle measurement more practical. We observe $T_2$ = (0.90 ± 0.05) ms at $T$ = 20 K and $B$ = 253 G (Fig. 1d). These coherence times are substantially longer than both those in previous reports in Ky5 ensembles ($T_2^*$ = 52 ns and $T_2$ = 24 us) [33], longer than those seen in non-isotopically purified diamond NV center ensembles ($T_2$ = 0.63 ms) [42], and comparable to those of 4H-SiC divacancies ($T_2$ = 1.3 ms) [28,43]. It is likely that this coherence time can be improved to several milliseconds in isotopically purified SiC or by using more sophisticated dynamical decoupling sequences [43,44].

To identify Ky5/L3, we compare *ab initio* density functional theory (DFT) simulations of the 3C-SiC neutral divacancy to experimental reconstructions of the Ky5/L3 hyperfine coupling tensors. We use a 512-atom simulation of 3C-SiC in VASP to compute the principal axes of the neutral divacancy's $^{29}$Si-IIa, $^{29}$Si-IIb, and $^{13}$C-I hyperfine tensors (Table 1). These simulations use the projector augmented wave approach with HSE06 functionals [45–47], which we expect to be accurate to within 20% or better ($\Delta v \approx$ 1.5 MHz for $^{29}$Si IIa and IIb). Single spins have less inhomogeneity, and hence tend to have larger $T_2^*$ coherence times, than ensembles. This increases the spectral resolution of our pulsed ODMR measurements, and we are able to detect shifts of less than 1 MHz.

We locate two single Ky5 defects where we observe low-field hyperfine splittings of 59 MHz and 9 MHz in pulsed ODMR measurements, which are near the expected $^{13}$C-I and $^{29}$Si-IIb resonances. On each defect, we repeat pulsed ODMR measurements across both $m_s = \pm 1$ branches at different magnetic field strengths and angles ($B$ = 10-250 G, $\theta$ = 0-80°). The varied fields hybridize the electronic and nuclear spin states in different ways, which manifest as small shifts in the hyperfine resonance locations. By simultaneously fitting each of these ODMR scans, we infer the full hyperfine tensors (Table 1) parameterized by $A_{xx}$, $A_{yy}$, $A_{zz}$, and angle $\theta$ (see Supplementary Information) [48]. The fit uncertainties indicate we are not sensitive to $A_{yy}$, so we set $A_{yy} = A_{xx}$ as an approximation. We also infer the $A_z$ value for the $^{29}$Si-IIa site from the previously-described ensemble Hahn echo measurement's electron spin echo envelope modulations (see Supplementary Information). Besides the hyperfine tensors, our computations also predict a ground state ZFS of $D$ = 1.32 GHz for the 3C neutral divacancy, which is consistent with the 1.336 GHz value we measure here to about the same accuracy as *ab initio* calculations in the 4H and 6H polytypes [25]. The excellent quantitative agreements between our experiments and the simulated hyperfine couplings and ZFSs are very strong evidence that Ky5/L3 is the neutral divacancy.

**RESONANT EXCITATION OF SiC DIVACANCIES**

The optical fine structure of a defect can enable initialization and readout protocols that have substantially higher fidelity than approaches using standard off-resonant initialization and readout. Because of solid-state strain inhomogeneity, using resonant excitation to observe the optical fine structure directly requires measuring a single defect, not an ensemble. We have isolated single divacancies in 3C-SiC, and to extend our studies to 4H-SiC, we prepare a sample from the same wafer as the previous 4H-SiC study where single divacancies were isolated [28,49]. The single divacancy in 3C-SiC and the (*hh*) and (*kk*) forms of divacancy in 4H-

SiC should each have a $^3E$ excited state because of their $C_{3v}$ symmetry and because each has six active electrons. We focus on these forms because a detailed theoretical treatment of this structure, developed in the context of diamond NV centers, is already available in two recent works [50,51].

To emphasize the importance of quantifying interactions in the excited state and their impact on the high-fidelity control of divacancies, we diagram the response of the six $^3E$ excited state sublevels to spin-orbit ($\lambda_z$), spin-spin ($D_{es}$, $\Delta_1$, $\Delta_2$), and transverse strain ($\delta_{perp}$) interactions in Fig. 2a. The effect of $\lambda_z$, $D_{es}$, and $\Delta_1$ is to split or shift certain sublevels, while the effect of $\Delta_2$ is to mix the $E_1$ and $E_2$ states with the $E_x$ and $E_y$ states. Transverse strain will alter the level energies further, and also cause states of differing spin to mix. This degrades the achievable readout fidelities since it increases the chance that optical excitation will induce the spin to flip. At transverse strains that are low relative to $\lambda_z$, and if $\Delta_2$ is small, the $E_x$ and $E_y$ levels will have an almost perfect $m_s = 0$ character. Thus, when either of these states is optically excited, the probability of a spin flip, $p$, is very small since spin mixing with states of $m_s = \pm 1$ character is almost negligible. Excitation will strongly project the ground state spin along the $m_s = 0$ basis state, and the emission of a photon signals that the spin was in the $m_s = 0$ state. This transition can be then repeatedly cycled to produce an average of $1/p$ photons before a spin flip occurs. For single-shot readout, both the number of photons extracted and the collection efficiencies must be as high as possible, and thus a precise understanding of the excited state is vital for entangling distant SiC defect spins [52–55].

To create single divacancies in 4H-SiC, we irradiate our sample with 2 MeV electrons at room temperature to a fluence of 5 x $10^{12}$ cm$^{-2}$, and anneal it at 745 °C in Ar gas for 30 minutes. Single divacancies are easily seen in spatial PL scans (see Supplementary Information). At temperatures below $T$ = 20 K, transitions between different ground and excited state sublevels become sharp and we can observe the optical fine structure in a photoluminescence excitation (PLE) measurement. In this measurement, an off-resonant laser pulse first polarizes the ground state spin into $m_s = 0$, and we detect the sideband PL as we scan the frequency of a tunable narrow-line laser. Optionally, we can apply continuous microwaves at a ground state $m_s = \pm 1$ magnetic resonance transition so that we can also observe transitions from those ground state sublevels. Fig. 2b shows the PLE spectrum of a single 4H-SiC (*kk*) divacancy measured at T = 8 K. Each of the six spin-allowed transitions are seen, and they are split by a transverse strain of $\delta_{perp}$ = 7 GHz. If no microwaves are applied, only the $E_x$ and $E_y$ transitions are seen (not shown), which implies that off-resonant light polarizes the spin to a high degree into $m_s$ = 0. This is also consistent with the strong polarization seen in single 3C-SiC divacancies using time-resolved PL measurements (Appendix A).

The strength of each interaction term in the excited-state Hamiltonian can be inferred by analyzing the relative PLE resonance frequencies as a function of transverse strain [56]. We perform PLE measurements on 13 (*hh*) and 10 (*kk*) divacancies, which we find at randomly distributed transverse strains, and use a Bayesian analysis (see Supplementary Information) to infer the spin-orbit and spin-spin interaction parameters intrinsic to both forms. The data for the (*kk*) divacancy and the model fit are shown in Fig. 2c. A similar plot for (*hh*) divacancies, along with raw PLE data taken on a single basally oriented (*kh*) divacancy, can be found in the Supplementary Information. The agreement of the data and model for both (*hh*) and (*kk*) forms is generally within the 10 MHz repeatability of our wavemeter. For the (*hh*) divacancies near 264.91 THz, we infer $\lambda_z$ = (3.538 ± 0.052) GHz, $D_{es}$ = (0.855 ± 0.017) GHz, $\Delta_1$ = (0.577 ± 0.019) GHz, and $\Delta_2 = (0.031^{+0.050}_{-0.031})$ GHz, while for the (*kk*) divacancies near 265.31 THz, we infer $\lambda_z$ = (6.090 ± 0.052) GHz, $D_{es}$ = (0.852 ± 0.012) GHz, $\Delta_1$ = (0.584 ± 0.012) GHz, and $\Delta_2 = (0.044^{+0.046}_{-0.044})$ GHz, where all values are quoted with a 95% credible interval. As mentioned, the same general $^3E$ structure is also present in NV centers, but the specific interaction strengths of SiC divacancies mean that spin mixing in the $E_x$ level is an order of magnitude weaker (see Supplementary Information). This is mostly a consequence of having a weaker mixing term ($\Delta_2$ ~ 40 MHz vs ~200 MHz) with a comparable spin-orbit strength (3.5-6 GHz vs 5.3 GHz) [56]. We also observe no ionization or spectral hopping of the transitions between scans [1,55,57]. These effects, along with photon losses, are hindrances in current diamond-based single-shot readout and entanglement efforts. The divacancy's intrinsic ability to emit at least as many photons per shot and to remain in the correct charge and resonance states over time will be a boon to achieving single-shot readout and faster entanglement success rates in SiC.

To investigate the 3C-SiC divacancy, we perform the same PLE measurement on single divacancies at T = 8 K near 270.95 THz. Fig. 2d shows the PLE resonances of different single 3C divacancies that experience transverse strains of $\delta_{perp}$ = 22-65 GHz. The splitting of the two excited state branches with strain follows the characteristic behavior expected for the $^3E$ excited state, and we infer $\lambda_z$ = (15.7 ± 2.5) GHz, $D_{es}$ = (2.0 ± 1.0) GHz, and $\Delta_1$, $\Delta_2$ < 1.0 GHz for the 3C-SiC divacancy's excited state. The linewidths we observe in this sample are nearly 2 GHz, which are significantly larger than the typical 100-200 MHz linewidths we observe in 4H-SiC divacancies. We suspect that the larger linewidths originate from the significantly higher 0.1-0.5 ppm nitrogen content in this 3C-SiC sample versus the 1 ppb content in our 4H-SiC sample. Given these linewidths, we cannot precisely infer the weaker spin-spin interaction terms, $\Delta_1$ and $\Delta_2$, which are critical for predicting spin mixing behavior. We note, however, that the spin-orbit coupling interaction is significantly stronger here, and if the values of $\Delta_1$ and $\Delta_2$ are similar to what we observed in the two forms of 4H-SiC divacancy, mixing should be suppressed in 3C-SiC by an additional order of magnitude compared to 4H-SiC (see Supplementary Information). One additional peculiarity of this sample is that the PLE peak intensities change only negligibly when microwaves are applied. We might observe this behavior for an unpolarized spin with short spin coherence, but the off-resonant ODMR contrast being comparable to divacancies in 4H-SiC, the long Hahn echo spin coherence time we measure, and the high polarization we infer from time-resolved PL (Appendix A) all suggest this is not the case. We surmise that a lower impurity sample will eliminate this behavior and will display linewidths comparable to 4H-SiC.

An optical transition's linewidth should be as narrow as possible so that two-photon interference can be observed, which is necessary for remote entanglement [58,59]. The lifetime-limited linewidths for each of the divacancies we measure are near 11 MHz, but the lowest linewidth we observe is about 80 MHz on a particular (*kk*) divacancy. Impurities in solids, including diamond, often broaden optical linewidths via temperature-independent charge fluctuations [60–62]. The typical 100-200 MHz linewidths we measure in our 4H-SiC sample are comparable to the 36-483 MHz linewidths of NV centers used in spin-photon and spin-spin entanglement experiments [59,63,64]. This is already promising, but it is important to establish whether the divacancy's optical linewidths are broadened here by intrinsic interactions with lattice phonons or by impurities that can be reduced in future efforts. To investigate, we measure the optical linewidths of each form of divacancy in both samples as a function of temperature (Fig. 3a). At temperatures above T = 20 K, the linewidth of each form follows a $T^5$ dependence [65,66], while at lower temperatures, the linewidths of each form approach an asymptote. Notably, the low-temperature linewidths are the same (~100 MHz) for different inequivalent divacancies in our 4H-SiC sample, where the nitrogen concentration is a few ppb, while divacancies in the 3C-SiC sample have about 20x larger linewidths (2 GHz), where the nitrogen concentration is about 0.5-1 ppm. The identical linewidths among inequivalent forms in the same 4H-SiC sample, similar broadening at high-temperatures, temperature-independence below 20 K, and prior reports of poor linewidths in nitrogen-rich diamond, are all consistent with the idea that linewidths we observe are sample-dependent rather than intrinsic [60,66–69]. These observations suggest that reducing impurities and annealing any residual crystal damage are promising routes towards achieving lifetime-limited linewidths in SiC divacancies [70,71].

We have so far probed the excited-state structure of divacancies in SiC, and our measurements predict that it can be used for high-fidelity polarization and readout. We now show explicitly that certain optical transitions are highly spin-dependent, and that they can be used for high-contrast detection of the ground state spin. The Rabi oscillations of a single 4H-SiC (*hh*) divacancy detected by resonantly exciting its $E_y$ transition are shown in Fig. 3b. Here we use an off-resonant laser pulse to prepare the spin and apply a variable-length microwave burst for spin rotation (as in Fig. 1c), but the spin is detected by applying a resonant laser pulse and monitoring the sideband PL. While standard off-resonant readout gives only a 9% contrast [28], the measurement here has a 94% readout contrast. This significant improvement originates from both the high ground state polarization induced by the divacancy optical cycle (consistent with Appendix A), and because the spin mixing of $E_y$ at this divacancy's transverse strain of $\delta_{perp}$ = 27 GHz is not excessive. In the future, a second laser can be used to improve the spin polarization closer to unity by driving the $E_{1,2}$ transitions and depleting the residual $m_s = \pm 1$ population. In addition, we expect that using ultra-pure layers without removing the substrate will produce lower strain ($\delta_{perp}$ < 3 GHz) divacancies with superior spin-mixing properties.

The fidelity of single-shot readout depends sensitively on the number of photons detected per measurement shot, and the ultimate fidelities that can be obtained in practice will depend on both the total number of photons emitted and the efficiency by which they can be collected. Improved collection efficiencies were recently demonstrated in SiC defects [31], and our earlier spin-mixing analysis suggested that the (*hh*) and (*kk*) divacancies should have a low spin-flip probability and, as a consequence, emit at least as many photons per shot as the NV center. Besides spin-mixing, however, electron-phonon interactions in the excited state [66] or other effects that are currently unknown to us could increase the spin-flip probability realized in experiment and hence degrade the total number of photons emitted per shot. To gain a relative measure of this probability, we apply an off-resonant pulse to polarize a single divacancy's spin, and then time-resolve the photons emitted when we drive an $E_{x,y}$ transition. As the spin is repeatedly excited, it has more chances to flip, and the spin-flip rate can be inferred from the PL decay rate. Fig. 3b shows the spin-flip rates of a few divacancies, which increase as the laser power is increased as the spin has more chances to flip per unit time, and eventually saturate because of the divacancies' finite optical lifetime. Even with the sub-optimal transverse strain of our current sample, the saturation rates are near 330 kHz, and within the 150-400 kHz saturation rates of NV centers already used for high-fidelity single shot readout [55]. Since the optical lifetimes of (*hh*) and (*kk*) divacancies are similar to the NV center [25,72], these data suggest the total number of photons emitted is similar. With enhanced collection efficiencies, then, it appears that the fundamental physics of spin-flips should be sufficient for high-fidelity single-shot readout of SiC divacancies.

**CONCLUSION**

We performed PL, PLE, and magnetic resonance experiments on single 3C- and 4H-SiC divacancies aimed at revealing fundamental aspects of their electronic structure and their optical and spin properties. We find that the divacancy in 3C-SiC has a Hahn echo electronic spin coherence time that are nearly one millisecond long. This is an exciting result since 3C-SiC can be heteroepitaxially grown as a thin film on Si, which provides a natural way to obtain suspended membranes that can be processed into photonic cavities. In both 3C-SiC and 4H-SiC, we revealed a highly spin-dependent optical fine structure that provides a pathway to high-fidelity initialization and single-shot readout, which are necessary for remote entanglement protocols. It is also clear that the materials-growth of 4H-SiC is already sufficient to produce the narrow optical linewidths required for photon-mediated entanglement, but that improvements in impurity control will likely lead to narrower transitions in both 3C-SiC and 4H-SiC. We hope that our findings will stimulate research into using SiC divacancy spin qubits as building blocks in future quantum networks.

**ACKNOWLEDGEMENTS**

We are grateful to Paolo Andrich for assistance in sample preparation, and Abram L. Falk and Brian B. Zhou for helpful discussions. This work was funded by ARO W911NF-15-2-0058, AFOSR FA9550-15-1-0029 and FA9550-14-1-0231, NSF MRSEC DMR-1420709, the DOE LDRD Program, Swedish Research Council (621-2014-5825 and 2016-04068), ÅForsk foundation (16-576), Carl-Trygger Stiftelse för Vetenskaplig Forskning (CTS 15:339), Knut and Alice Wallenberg Foundation (KAW 2013.0300), and JSPS KAKENHI B 26286047.

**APPENDIX A: RATE EQUATION MODELING OF 3C-SIC DIVACANCY OPTICAL CYCLE**

It is empirically clear that off-resonant cycling of divacancies can both polarize and detect their electronic spin states [12,15], but prior works have not focused on the details of the divacancy's optical cycle that underlie these functionalities. Based on their common $C_{3v}$ symmetry and six active electrons, the (*hh*) and (*kk*) divacancies in 4H-SiC, the divacancy in 3C-SiC, and the nitrogen-vacancy center in diamond should share the same electronic orbital levels but with different relative energy positions and different transition rates between levels. Besides the $^3A_2$ ground state and the $^3E$ excited state, which we focused on in the main text, the $^1A_1$ and $^1E$ spin singlet levels also exist [11,73]. The specific transition rates between orbital and spin levels are responsible for how much spin polarization can be generated, the speed of the polarization and readout process, and the change in PL when the spin populates an $m_s = \pm 1$ sublevel versus $m_s = 0$. These are important fundamental questions for future experiments, and we can use time-resolved PL experiments on single spins to establish a simplified model of the 3C-SiC divacancy's optical cycle.

We propose a five-level model of the divacancy's spin and orbital levels in Fig. 4a. In this model, the ground and excited states are treated as orbital singlets with one level corresponding to $m_s$ = 0 and another corresponding to the $m_s$ = ±1 within each orbital. A fifth level represents one or more spin singlet states, and transitions into and out of this level are treated as non-radiative. For the divacancy's spin to both polarize under repeated optical excitation and generate a change in fluorescence based on its spin state, at least one transition should be spin-selective and non-radiative. One possible mechanism, in analogy with the NV center, is that the $m_s$ = ±1 sublevels of the $^3E$ state undergo a non-radiative intersystem crossing (ISC) to a spin singlet state via a non-axial spin-orbit interaction [52,74].

To explicitly probe spin-dependent decay pathways from the excited state, we time-resolve the PL emitted when a short laser pulse is applied after the ground state spin is prepared in $m_s$ = 0 and $m_s$ = ±1. In this experiment, a 1 μs long 975 nm off-resonant pulse polarizes the spin into $m_s$ = 0, a 1 μs delay allows the defect to relax back to its ground state, and a sub-nanosecond 920 nm pulse from a mode-locked Ti:Sapphire laser excites the spin [72,75,76]. The PL induced by this last pulse is time-resolved in Fig. 4b when no microwaves are applied, and when a π-pulse on the $m_s$ = -1 transition is applied just before the last pulse. The PL follows a biexponential decay in both traces, and the relative amplitudes of the two lifetime components track the proportion of $m_s$ = 0 and $m_s$ = ±1 at intermediate spin rotation angles [77,78]. The shorter lifetime is indicative of a non-radiative decay out of the excited state, and the dependence on the ground state spin indicates it is spin-selective. This result is analogous to similar experiments in NV centers, and is solid evidence that their optical cycles function in a similar way [72]. The solid lines are fits we perform on both datasets simultaneously (see Supplementary Information) and reveal the two lifetimes as $\tau_0$ = (18.7 ± 0.3) ns and $\tau_1$ = (15.7 ± 0.3) ns. The difference in the two lifetimes suggests that the excited state to singlet transitions differ by (10 ± 2) MHz, depending on whether the spin is in $m_s$ = ±1 or $m_s$ = 0. This measurement also indicates that the ground state spin polarization is quite high. By computing the relative fraction of the two lifetime components, we infer a $96.5^{+4}_{-5}$% spin polarization in the ground state. This is close to the high polarization required for a 94% Rabi oscillation contrast observed in a 4H-SiC divacancy in Fig. 3b, which illustrates another similarity between divacancies in different polytypes.

To infer the remaining rates, we perform a second experiment where we prepare the spin in either $m_s = 0$ or $m_s = -1$ using the same CW laser pulse and microwave pulses as the previous experiment, but time-resolve the PL emitted under a second CW laser pulse. The PL dynamics under continuous excitation are more complicated than biexponential decay since the PL no longer depends on only a subset of the defect's transition rates. To control for overfitting, we perform the experiment at eleven laser powers between 0.35 and 2.87 mW and perform a fit of the rate-equation model to all eleven pairs of datasets simultaneously. This excitation range covers below and above the intensity at which the defect's PL saturates (see Supplementary Information). During the fitting process, the ISC rates and polarization are constrained by the biexponential decay experiment, and the relative optical pumping strengths are fixed based on the measured laser intensities. Allowing for decay from the singlet states to $m_s = \pm 1$ levels in the ground state did not improve the fit significantly, so this term was set to zero. Four experimental runs where the spin is prepared in either $m_s = 0$ or $m_s = -1$ are shown in Fig. 4d, where the solid lines in each subplot are the output of the global fitting procedure.

The amount of similarity between the orbital and spin-selective dynamics of SiC divacancies and diamond NV centers has been an open question, and our model's quantitative reproduction of the observed PL dynamics is good evidence that the special optical cycle seen in NV centers is also present in SiC divacancies. The biexponential decay experiment explicitly shows that a spin-selective and non-radiative transition exists out of the excited state, which is likely the reason why the divacancy's spin can be polarized and detected using off-resonant laser light. This transition is only weakly spin-selective, however, and is about 7x slower than the corresponding NV center rate. The singlet decay rate we infer for divacancies is also about 20x faster than that of NV centers [72], and consistent with recent qualitative ensemble measurements [79]. Together, this means that a spin in the $m_s = \pm 1$ state will undergo a non-radiative transition less often, and that when it does it quickly decays to the ground state where it can be excited again and emit photons. This likely explains the weaker 7.5% off-resonant readout contrast (Fig. 1c) of 3C-SiC divacancies relative to the nearly 30% seen in NV centers, and may also explain the 9-15% contrasts seen in single 4H-SiC divacancies [28]. We also found that allowing the $m_s = 0$ level in the excited state to undergo an ISC to the singlet noticeably improved the quality of the fit, but a transverse spin-orbit interaction should only couple to states with $m_s = \pm 1$ spin [80]. This may be partially explained by this sample's high strain, since it will partially mix the excited state sublevels and violate our model's assumption of sublevels with pure spin character. As mentioned, the fit is constrained by the biexponential decay experiment, so the model suggests the optical lifetime of 23 ns is shorted to an apparent value of 18.7 ns because of this additional non-radiative ISC decay. In the future, measuring all six excited state branching ratios using resonant spectroscopy on a low-strain sample will give the clearest insight into the details of the divacancy's ISC [80].

**REFERENCES**

[1]  H. Bernien, B. Hensen, W. Pfaff, G. Koolstra, M. S. Blok, L. Robledo, T. H. Taminiau, M. Markham, D. J. Twitchen, L. Childress, and R. Hanson, "Heralded Entanglement Between Solid-State Qubits Separated by Three Metres," Nature **497**, 86 (2013).

[2]  W. Pfaff, B. J. Hensen, H. Bernien, S. B. van Dam, M. S. Blok, T. H. Taminiau, M. J. Tiggelman, R. N. Schouten, M. Markham, D. J. Twitchen, and R. Hanson, "Unconditional Quantum Teleportation Between Distant Solid-State Quantum Bits," Science **345**, 532 (2014).

[3]  B. Hensen, H. Bernien, A. E. Dréau, A. Reiserer, N. Kalb, M. S. Blok, J. Ruitenberg, R. F. L. Vermeulen, R. N. Schouten, C. Abellán, W. Amaya, V. Pruneri, M. W. Mitchell, M. Markham, D. J. Twitchen, D. Elkouss, S. Wehner, T. H. Taminiau, and R. Hanson, "Loophole-Free Bell Inequality Violation Using Electron Spins Separated by 1.3

Kilometres," Nature **526**, 682 (2015).

[4] L. J. Rogers, K. D. Jahnke, M. H. Metsch, A. Sipahigil, J. M. Binder, T. Teraji, H. Sumiya, J. Isoya, M. D. Lukin, P. Hemmer, and F. Jelezko, "All-Optical Initialization, Readout, and Coherent Preparation of Single Silicon-Vacancy Spins in Diamond," Phys. Rev. Lett. **113**, 263602 (2014).

[5] J. R. Rabeau, Y. L. Chin, S. Prawer, F. Jelezko, T. Gaebel, and J. Wrachtrup, "Fabrication of Single Nickel-Nitrogen Defects in Diamond by Chemical Vapor Deposition," Appl. Phys. Lett. **86**, 131926 (2005).

[6] R. Kolesov, K. Xia, R. Reuter, R. Stöhr, A. Zappe, J. Meijer, P. R. Hemmer, and J. Wrachtrup, "Optical Detection of a Single Rare-Earth Ion in a Crystal," Nat. Commun. **3**, 1029 (2012).

[7] P. Siyushev, K. Xia, R. Reuter, M. Jamali, N. Yang, C. Duan, N. Kukharchyk, A. D. Wieck, J. Wrachtrup, N. Zhao, and R. Kolesov, "Coherent Properties of Single Rare-Earth Spin Qubits," Nat. Commun. **5**, 3895 (2014).

[8] W. F. Koehl, B. Diler, S. J. Whiteley, A. Bourassa, N. T. Son, E. Janzén, and D. D. Awschalom, "Resonant optical spectroscopy and coherent control of Cr4+ spin ensembles in SiC and GaN," Phys. Rev. B **95**, 35207 (2017).

[9] J. R. Weber, W. F. Koehl, J. B. Varley, A. Janotti, B. B. Buckley, C. G. Van de Walle, and D. D. Awschalom, "Quantum Computing with Defects," Proc. Natl. Acad. Sci. U. S. A. **107**, 8513 (2010).

[10] P. G. Baranov, A. P. Bundakova, A. A. Soltamova, S. B. Orlinskii, I. V. Borovykh, R. Zondervan, R. Verberk, and J. Schmidt, "Silicon Vacancy in SiC as a Promising Quantum System for Single-Defect and Single-Photon Spectroscopy," Phys. Rev. B **83**, 125203 (2011).

[11] A. Gali, "Time-Dependent Density Functional Study on the Excitation Spectrum of Point Defects in Semiconductors," Phys. Status Solidi **248**, 1337 (2011).

[12] W. F. Koehl, B. B. Buckley, F. J. Heremans, G. Calusine, and D. D. Awschalom, "Room Temperature Coherent Control of Defect Spin Qubits in Silicon Carbide," Nature **479**, 84 (2011).

[13] H. Kraus, V. a. Soltamov, D. Riedel, S. Väth, F. Fuchs, A. Sperlich, P. G. Baranov, V. Dyakonov, and G. V. Astakhov, "Room-Temperature Quantum Microwave Emitters Based on Spin Defects in Silicon Carbide," Nat. Phys. **10**, 157 (2013).

[14] S. Castelletto, B. C. Johnson, V. Ivády, N. Stavrias, T. Umeda, A. Gali, and T. Ohshima, "A Silicon Carbide Room-Temperature Single-Photon Source," Nat. Mater. **12**, 151 (2013).

[15] P. G. Baranov, I. V. Il'in, E. N. Mokhov, M. V. Muzafarova, S. B. Orlinskii, and J. Schmidt, "EPR Identification of the Triplet Ground State and Photoinduced Population Inversion for a Si-C Divacancy in Silicon Carbide," J. Exp. Theor. Phys. Lett. **82**, 441 (2005).

[16] N. T. Son, P. Carlsson, J. Ul Hassan, E. Janzén, T. Umeda, J. Isoya, A. Gali, M. Bockstedte, N. Morishita, T. Ohshima, and H. Itoh, "Divacancy in 4H-SiC," Phys. Rev. Lett. **96**, 55501 (2006).

[17] S. E. Saddow and A. Agarwal, "Advances in Silicon Carbide Processing and Applications" (Artech House Publishers, London, 2004).

[18] A. Powell, J. Sumakeris, Y. Khlebnikov, M. Paisley, R. Leonard, E. Deyneka, S. Gangwal, J. Ambati, V. Tsevtkov, J. Seaman, A. McClure, J. Guo, M. Dudly, E. Balkas, and V. Balakrishna, "Bulk Growth of Large Area SiC Crystals," Mater. Sci. Forum **858**, 5 (2016).

[19] C. A. Zorman, A. J. Fleischman, A. S. Dewa, M. Mehregany, C. Jacob, S. Nishino, and P. Pirouz, "Epitaxial Growth of 3C-SiC Films on 4 in. Diam (100) Silicon Wafers by Atmospheric Pressure Chemical Vapor Deposition," J. Appl. Phys. **78**, 5136 (1995).

[20] Y. T. Yang, K. L. Ekinci, X. M. H. Huang, L. M. Schiavone, M. L. Roukes, C. A. Zorman, and M. Mehregany, "Monocrystalline Silicon Carbide Nanoelectromechanical Systems," Appl. Phys. Lett. **78**, 162 (2001).

[21] G. Calusine, A. Politi, and D. D. Awschalom, "Silicon Carbide Photonic Crystal Cavities with Integrated Color Centers," Appl. Phys. Lett. **105**, 11123 (2014).

[22] J. C. Lee, A. P. Magyar, D. O. Bracher, I. Aharonovich, and E. L. Hu, "Fabrication of Thin Diamond Membranes for Photonic Applications," Diam. Relat. Mater. **33**, 45 (2013).

[23] J. L. O'Brien, A. Furusawa, and J. Vučković, "Photonic Quantum Technologies," Nat. Photonics **3**, 687 (2010).

[24] I. Aharonovich, A. D. Greentree, and S. Prawer, "Diamond Photonics," Nat. Photonics **5**,

397 (2011).

[25] A. L. Falk, P. V. Klimov, B. B. Buckley, V. Ivády, I. A. Abrikosov, G. Calusine, W. F. Koehl, Á. Gali, and D. D. Awschalom, "Electrically and Mechanically Tunable Electron Spins in Silicon Carbide Color Centers," Phys. Rev. Lett. **112**, 187601 (2014).

[26] P. V. Klimov, A. L. Falk, B. B. Buckley, and D. D. Awschalom, "Electrically Driven Spin Resonance in Silicon Carbide Color Centers," Phys. Rev. Lett. **112**, 87601 (2014).

[27] O. V Zwier, D. O'Shea, A. R. Onur, and C. H. van der Wal, "All-Optical Coherent Population Trapping with Defect Spin Ensembles in Silicon Carbide," Sci. Rep. **5**, 10931 (2015).

[28] D. J. Christle, A. L. Falk, P. Andrich, P. V Klimov, J. Ul Hassan, N. T. Son, E. Janzén, T. Ohshima, and D. D. Awschalom, "Isolated Electron Spins in Silicon Carbide with Millisecond-Coherence Times," Nat. Mater. **14**, 160 (2015).

[29] H. J. Kimble, "The Quantum Internet," Nature **453**, 1023 (2008).

[30] W. B. Gao, A. Imamoglu, H. Bernien, and R. Hanson, "Coherent Manipulation, Measurement and Entanglement of Individual Solid-State Spins Using Optical Fields," Nat. Photonics **9**, 363 (2015).

[31] M. Widmann, S.-Y. Lee, T. Rendler, N. T. Son, H. Fedder, S. Paik, L.-P. Yang, N. Zhao, S. Yang, I. Booker, A. Denisenko, M. Jamali, S. A. Momenzadeh, I. Gerhardt, T. Ohshima, A. Gali, E. Janzén, and J. Wrachtrup, "Coherent Control of Single Spins in Silicon Carbide at Room Temperature," Nat. Mater. **14**, 164 (2014).

[32] J. Y. Lee, X. Lu, and Q. Lin, "High-Q Silicon Carbide Photonic-Crystal Cavities," Appl. Phys. Lett. **106**, 41106 (2015).

[33] A. L. Falk, B. B. Buckley, G. Calusine, W. F. Koehl, V. V Dobrovitski, A. Politi, C. A. Zorman, P. X.-L. Feng, D. D. Awschalom, V. Viatcheslav, A. Politi, C. A. Zorman, X. Feng, and D. D. Awschalom, "Polytype Control of Spin Qubits in Silicon Carbide," Nat. Commun. **4**, 1819 (2013).

[34] J. Cardenas, M. Zhang, C. T. Phare, S. Y. Shah, C. B. Poitras, B. Guha, and M. Lipson, "High Q SiC Microresonators," Opt. Express **21**, 16882 (2013).

[35] M. Radulaski, T. M. Babinec, K. Müller, K. G. Lagoudakis, J. L. Zhang, S. Buckley, Y. A. Kelaita, K. Alassaad, G. Ferro, and J. Vučković, "Visible Photoluminescence from Cubic

(3C) Silicon Carbide Microdisks Coupled to High Quality Whispering Gallery Modes," ACS Photonics **2**, 14 (2015).

[36] B.-S. Song, S. Yamada, T. Asano, and S. Noda, "Demonstration of Two-Dimensional Photonic Crystals Based on Silicon Carbide," Opt. Express **19**, 11084 (2011).

[37] V. Y. Bratus, R. S. Melnik, S. M. Okulov, V. N. Rodionov, B. D. Shanina, and M. I. Smoliy, "A New Spin One Defect in Cubic SiC," Physica B **404**, 4739 (2009).

[38] N. T. Son, E. Sörman, W. M. Chen, C. Hallin, O. Kordina, B. Monemar, E. Janzén, and J. L. Lindström, "Optically Detected Magnetic Resonance Studies of Defects in Electron-Irradiated 3C SiC Layers," Phys. Rev. B **55**, 2863 (1997).

[39] L. Gordon, A. Janotti, and C. G. Van de Walle, "Defects as Qubits in 3C- and 4H-SiC," Phys. Rev. B **92**, 45208 (2015).

[40] V. Jokubavicius, G. R. Yazdi, R. Liljedahl, I. G. Ivanov, J. Sun, X. Liu, P. Schuh, M. Wilhelm, P. Wellmann, R. Yakimova, and M. Syväjärvi, "Single Domain 3C-SiC Growth on Off-Oriented 4H-SiC Substrates," Cryst. Growth Des. **15**, 2940 (2015).

[41] H. J. Kimble, M. Dagenais, and L. Mandel, "Multiatom and Transit-Time Effects on Photon-Correlation Measurements in Resonance Fluorescence," Phys. Rev. A **18**, 201 (1978).

[42] P. L. Stanwix, L. M. Pham, J. R. Maze, D. Le Sage, T. K. Yeung, P. Cappellaro, P. R. Hemmer, A. Yacoby, M. D. Lukin, and R. L. Walsworth, "Coherence of Nitrogen-Vacancy Electronic Spin Ensembles in Diamond," Phys. Rev. B **82**, 201201 (2010).

[43] H. Seo, A. L. Falk, P. V. Klimov, K. C. Miao, G. Galli, and D. D. Awschalom, "Quantum Decoherence Dynamics of Divacancy Spins in Silicon Carbide," Nat. Commun. **7**, 12935 (2016).

[44] L. P. Yang, C. Burk, M. Widmann, S. Y. Lee, J. Wrachtrup, and N. Zhao, "Electron Spin Decoherence in Silicon Carbide Nuclear Spin Bath," Phys. Rev. B **90**, 241203 (2014).

[45] P. E. Blöchl, "Projector Augmented-Wave Method," Phys. Rev. B **50**, 17953 (1994).

[46] G. Kresse and J. Furthmüller, "Efficient Iterative Schemes for Ab Initio Total-Energy Calculations Using a Plane-Wave Basis Set," Phys. Rev. B **54**, 11169 (1996).

[47] J. P. Perdew, J. A. Chevary, S. H. Vosko, K. A. Jackson, M. R. Pederson, D. J. Singh, and C.

Fiolhais, "Atoms, Molecules, Solids, and Surfaces: Applications of the Generalized Gradient Approximation for Exchange and Correlation," Phys. Rev. B **46**, 6671 (1992).

[48] V. Ivády, K. Szász, A. L. Falk, P. V. Klimov, D. J. Christle, E. Janzén, I. A. Abrikosov, D. D. Awschalom, and A. Gali, "Theoretical Model of Dynamic Spin Polarization of Nuclei Coupled to Paramagnetic Point Defects in Diamond and Silicon Carbide," Phys. Rev. B **92**, 115206 (2015).

[49] J. Hassan, J. P. Bergman, A. Henry, and E. Janzén, "On-Axis Homoepitaxial Growth on Si-Face 4H-SiC Substrates," J. Cryst. Growth **310**, 4424 (2008).

[50] J. R. Maze, A. Gali, E. Togan, Y. Chu, A. Trifonov, E. Kaxiras, and M. D. Lukin, "Properties of Nitrogen-Vacancy Centers in Diamond: The Group Theoretic Approach," New J. Phys. **13**, 25025 (2011).

[51] M. W. Doherty, N. B. Manson, P. Delaney, and L. C. L. Hollenberg, "The Negatively Charged Nitrogen-Vacancy Centre in Diamond: The Electronic Solution," New J. Phys. **13**, 25019 (2011).

[52] N. B. Manson, J. P. Harrison, and M. J. Sellars, "Nitrogen-Vacancy Center in Diamond: Model of the Electronic Structure and Associated Dynamics," Phys. Rev. B **74**, 104303 (2006).

[53] P. Tamarat, N. B. Manson, J. P. Harrison, R. L. McMurtrie, A. Nizovtsev, C. Santori, R. G. Beausoleil, P. Neumann, T. Gaebel, F. Jelezko, P. Hemmer, and J. Wrachtrup, "Spin-Flip and Spin-Conserving Optical Transitions of the Nitrogen-Vacancy Centre in Diamond," New J. Phys. **10**, 45004 (2008).

[54] B. Naydenov, R. Kolesov, A. Batalov, J. Meijer, S. Pezzagna, D. Rogalla, F. Jelezko, and J. Wrachtrup, "Engineering Single Photon Emitters by Ion Implantation in Diamond," Appl. Phys. Lett. **95**, 181109 (2009).

[55] L. Robledo, L. Childress, H. Bernien, B. Hensen, P. F. Alkemade, and R. Hanson, "High-Fidelity Projective Read-Out of a Solid-State Spin Quantum Register," Nature **477**, 574 (2011).

[56] A. Batalov, V. Jacques, F. Kaiser, P. Siyushev, P. Neumann, L. J. Rogers, R. L. McMurtrie, N. B. Manson, F. Jelezko, and J. Wrachtrup, "Low Temperature Studies of the Excited-State

Structure of Negatively Charged Nitrogen-Vacancy Color Centers in Diamond," Phys. Rev. Lett. **102**, 195506 (2009).

[57] V. M. Acosta, C. Santori, A. Faraon, Z. Huang, K. M. C. Fu, A. Stacey, D. A. Simpson, K. Ganesan, S. Tomljenovic-Hanic, A. D. Greentree, S. Prawer, and R. G. Beausoleil, "Dynamic Stabilization of the Optical Resonances of Single Nitrogen-Vacancy Centers in Diamond," Phys. Rev. Lett. **108**, 206401 (2012).

[58] R. Lettow, Y. L. A. Rezus, A. Renn, G. Zumofen, E. Ikonen, S. Götzinger, and V. Sandoghdar, "Quantum Interference of Tunably Indistinguishable Photons from Remote Organic Molecules," Phys. Rev. Lett. **104**, 123605 (2010).

[59] H. Bernien, L. Childress, L. Robledo, M. Markham, D. Twitchen, and R. Hanson, "Two-Photon Quantum Interference From Separate Nitrogen Vacancy Centers in Diamond," Phys. Rev. Lett. **108**, 43604 (2012).

[60] D. E. McCumber and M. D. Sturge, "Linewidth and Temperature Shift of the R Lines in Ruby," J. Appl. Phys. **34**, 1682 (1963).

[61] W. E. Moerner, M. Orrit, and U. P. Wild, "Single-Molecule Detection, Imaging and Spectroscopy" (1996).

[62] J. Wolters, N. Sadzak, A. W. Schell, T. Schröder, and O. Benson, "Measurement of the Ultrafast Spectral Diffusion of the Optical Transition of Nitrogen Vacancy Centers in Nano-Size Diamond Using Correlation Interferometry," Phys. Rev. Lett. **110**, 27401 (2013).

[63] E. Togan, Y. Chu, a S. Trifonov, L. Jiang, J. Maze, L. Childress, M. V. G. Dutt, a S. Sørensen, P. R. Hemmer, a S. Zibrov, and M. D. Lukin, "Quantum Entanglement Between an Optical Photon and a Solid-State Spin Qubit," Nature **466**, 730 (2010).

[64] B. B. Buckley, G. D. Fuchs, L. C. Bassett, and D. D. Awschalom, "Spin-light Coherence for Single-Spin Measurement and Control in Diamond," Science **330**, 1212 (2010).

[65] M. B. Walker, "A T5 Spin-Lattice Relaxation Rate for Non-Kramers Ions," Can. J. Phys. **46**, 1347 (1968).

[66] K. M. C. Fu, C. Santori, P. E. Barclay, L. J. Rogers, N. B. Manson, and R. G. Beausoleil, "Observation of the Dynamic Jahn-Teller Effect in the Excited States of Nitrogen-Vacancy

Centers in Diamond," Phys. Rev. Lett. **103**, 256404 (2009).

[67] D. A. Wiersma, "Coherent Optical Transient Studies of Dephasing and Relaxation in Electronic Transitions of Large Molecules in the Condensed Phase," Adv. Chem. Phys. **47**, 421 (1981).

[68] F. Jelezko, I. Popa, A. Gruber, C. Tietz, J. Wrachtrup, A. Nizovtsev, and S. Kilin, "Single Spin States in a Defect Center Resolved by Optical Spectroscopy," Appl. Phys. Lett. **81**, 2160 (2002).

[69] P. Tamarat, T. Gaebel, J. R. Rabeau, M. Khan, A. D. Greentree, H. Wilson, L. C. L. Hollenberg, S. Prawer, P. Hemmer, F. Jelezko, and J. Wrachtrup, "Stark Shift Control of Single Optical Centers in Diamond," Phys. Rev. Lett. **97**, 83002 (2006).

[70] Y. Chu, N. P. De Leon, B. J. Shields, B. Hausmann, R. Evans, E. Togan, M. J. Burek, M. Markham, A. Stacey, A. S. Zibrov, A. Yacoby, D. J. Twitchen, M. Loncar, H. Park, P. Maletinsky, and M. D. Lukin, "Coherent Optical Transitions in Implanted Nitrogen Vacancy Centers," Nano Lett. **14**, 1982 (2014).

[71] Y. Chen, P. S. Salter, S. Knauer, L. Weng, A. C. Frangeskou, C. J. Stephen, S. N. Ishmael, P. R. Dolan, S. Johnson, B. L. Green, G. W. Morley, M. E. Newton, J. G. Rarity, M. J. Booth, and J. M. Smith, "Laser Writing of Coherent Colour Centres in Diamond," Nat. Photonics **11**, 77 (2016).

[72] L. Robledo, H. Bernien, T. Van Der Sar, and R. Hanson, "Spin Dynamics in the Optical Cycle of Single Nitrogen-Vacancy Centres in Diamond," New J. Phys. **13**, 25013 (2011).

[73] A. Gali, A. Gällström, N. T. Son, and E. Janzén, "Theory of Neutral Divacancy in SiC: A Defect for Spintronics," Mater. Sci. Forum **645–648**, 395 (2010).

[74] A. Lenef and S. C. Rand, "Electronic Structure of the N-V Center in Diamond: Theory," Phys. Rev. B **53**, 13441 (1996).

[75] D. M. Toyli, D. J. Christle, A. Alkauskas, B. B. Buckley, C. G. Van de Walle, and D. D. Awschalom, "Measurement and Control of Single Nitrogen-Vacancy Center Spins Above 600 K," Phys. Rev. X **2**, 31001 (2012).

[76] P. V Klimov, A. L. Falk, D. J. Christle, V. V Dobrovitski, and D. D. Awschalom, "Quantum Entanglement at Ambient Conditions in a Macroscopic Solid-State Spin Ensemble," Sci.

Adv. **1**, e1501015 (2015).

[77] N. B. Manson and R. L. McMurtrie, "Issues Concerning the Nitrogen-Vacancy Center in Diamond," J. Lumin. **127**, 98 (2007).

[78] A. Batalov, C. Zierl, T. Gaebel, P. Neumann, I. Y. Chan, G. Balasubramanian, P. R. Hemmer, F. Jelezko, and J. Wrachtrup, "Temporal Coherence of Photons Emitted by Single Nitrogen-Vacancy Defect Centers in Diamond Using Optical Rabi-Oscillations," Phys. Rev. Lett. **100**, 77401 (2008).

[79] G. Calusine, A. Politi, and D. D. Awschalom, "Cavity-Enhanced Measurements of Defect Spins in Silicon Carbide," Phys. Rev. Appl. **6**, 14019 (2016).

[80] M. L. Goldman, A. Sipahigil, M. W. Doherty, N. Y. Yao, S. D. Bennett, M. Markham, D. J. Twitchen, N. B. Manson, A. Kubanek, and M. D. Lukin, "Phonon-Induced Population Dynamics and Intersystem Crossing in Nitrogen-Vacancy Centers," Phys. Rev. Lett. **114**, 145502 (2015).

| Nucleus Site | $A_{xx}$ (MHz) | $A_{yy}$ (MHz) | $A_{zz}$ (MHz) | Θ (deg.) | $A_z$ (MHz) |
|---|---|---|---|---|---|
| $^{13}C_I$ | 51.3 | 52.0 | 122.2 | 72.6 | 61.0 |
|  | (49.5 ± 4.5) | (49.5 ± 4.5) | (108.5 ± 3.6) | (72.3 ± 4.3) | (57.6 ± 1.3) |
| $^{29}Si_{IIa}$ | 9.1 | 9.9 | 7.7 | 68.5 | 8.9 |
|  | (8.7 ± 1.0) | (8.7 ± 1.0) | (9.5 ± 1.0) | (47 ± 34) | (9.1 ± 0.2) |
| $^{29}Si_{IIb}$ | 11.4 | 11.4 | 11.8 | 50.1 | 11.6 |
|  | - | - | - | - | (12.4 ± 1.2) |

**TABLE 1.** Theoretical simulations and experimental reconstructions of $^{13}$C and $^{29}$Si hyperfine tensors. The experimental values are reported in parentheses with a 95% probability interval. The large uncertainty on θ for $^{29}Si_{IIa}$ is a consequence of the fact that the tensor is nearly isotropic within the MHz resolution of our experiment.

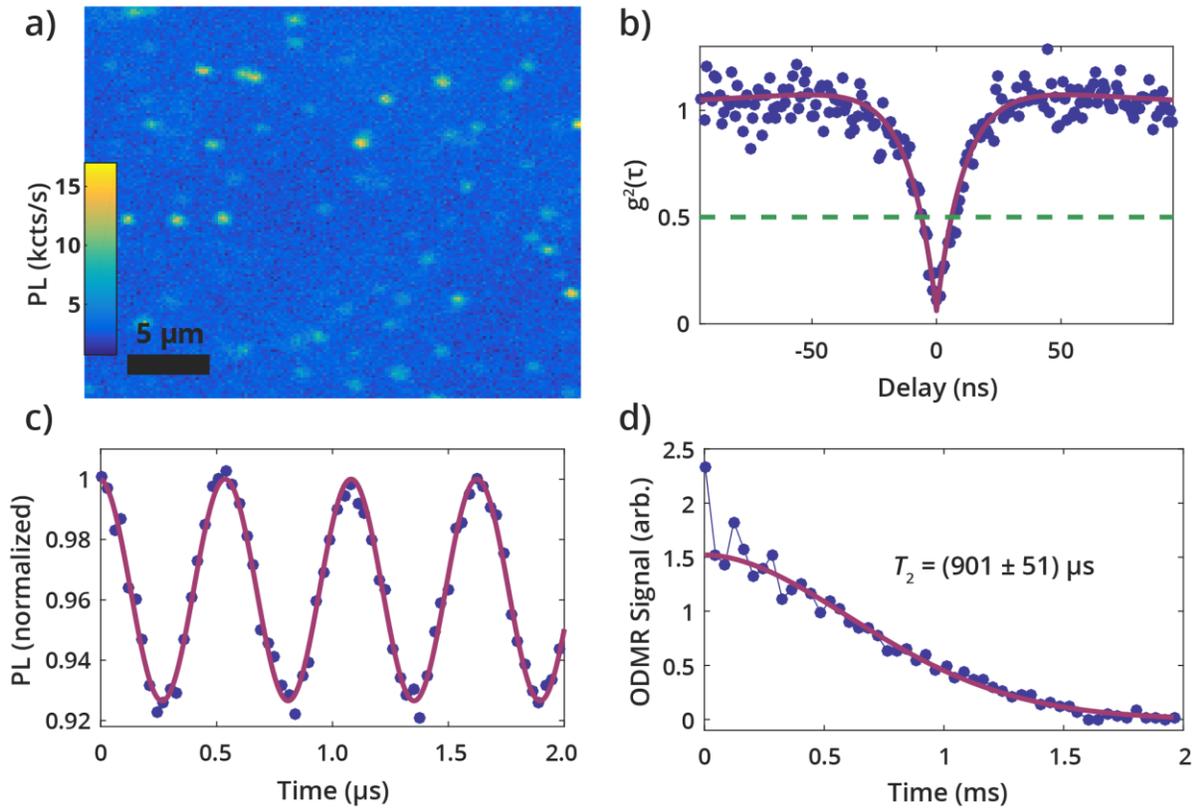

**FIG. 1. Isolation and control of single spins in 3C-SiC.** (a) A typical two-dimensional photoluminescence scan showing single defects as isolated luminescent spots in 3C-SiC. (b) Photon antibunching observed in the background-corrected photon correlation measurement taken an isolated photoluminescence spot. The green dashed line indicates the 0.5 threshold below which the emission is consistent with a single quantum emitter. Our analysis indicates $g^2(\tau = 0) = (0.055^{+0.044}_{-0.047})$ at the 95% probability level, which confirms we have isolated a single defect. (c) A measurement of Rabi oscillations recorded on the same defect, which shows an approximate off-resonant readout contrast of 7.5%. (*D*) Hahn echo measurement performed on an ensemble of Ky5 defects in a second sample at T = 20 K and B = 253 G showing $T_2$ = (901 ± 51) µs.

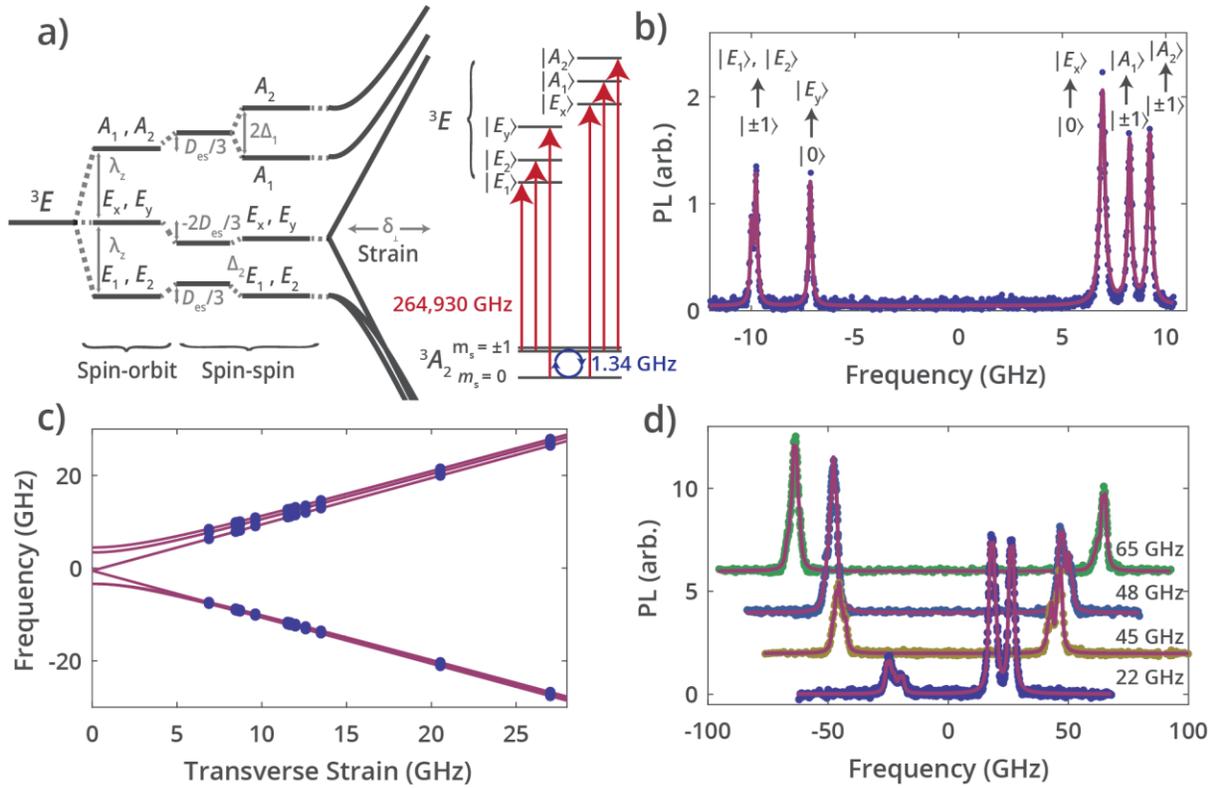

**FIG. 2 Excited-state fine structure of 4H and 3C-SiC divacancies.** (a) Level diagram of the $^3E$ excited state, showing the effect of axial spin-orbit coupling, the spin-spin interactions, and transverse strain. (b) Photoluminescence excitation measurements taken on a single 4H-SiC (*hh*) divacancy at T = 8 K under continuous microwave driving. (c) Individual resonances recorded at T = 8 K on separate (*hh*) divacancies that experience different magnitudes of transverse strain. The solid lines are the Hamiltonian energies at the best fit value. (d) Photoluminescence excitation measurements taken T = 8 K on single 3C-SiC divacancies that experience different transverse strains. The estimated magnitude of the transverse strain is labeled in units of GHz on each trace.

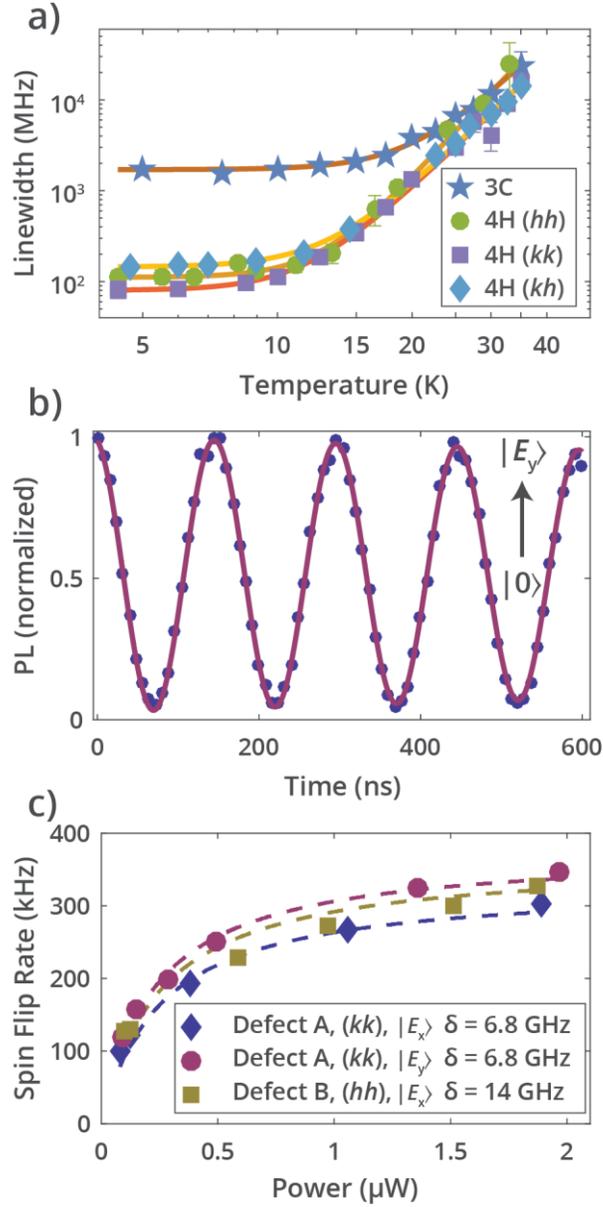

**FIG. 3. Characterization of divacancy optical transition properties.** (a) Temperature dependence of divacancy optical linewidths (b) Single spin Rabi oscillations recorded using the $E_y$ transition of a single (*hh*) divacancy at T = 8 K with a transverse strain of $\delta_{perp}$ = 27 GHz. The measurement gives a background-corrected readout contrast of 94%, consistent with a high degree of spin polarization. (c) The spin-flip rate as a function of resonant laser power measured on a different $E_{x,y}$ transitions on different single 4H-SiC divacancies.

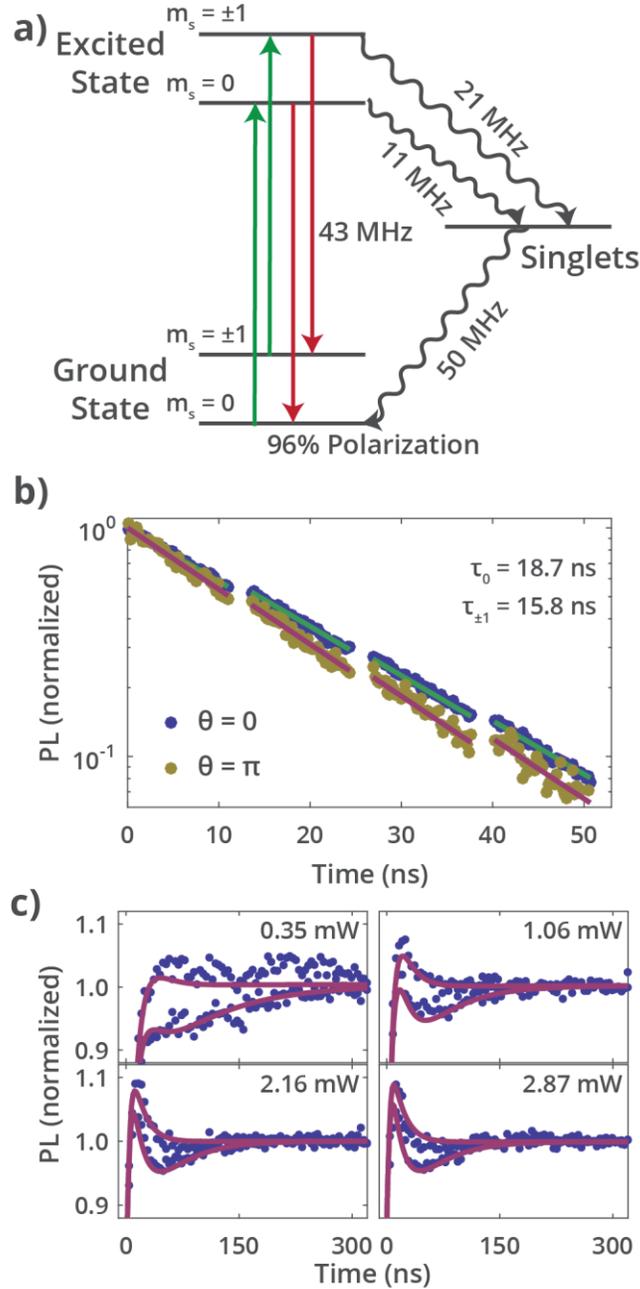

**FIG. 4. Dynamical model of the 3C-SiC divacancy.** (a) Diagram of the levels and major rates in the five-level rate-equation model. The transition rates and ground state spin polarization are inferred from the global fit described in the main text. (b) Normalized log plot of the fluorescence recorded on a single 3C-SiC divacancy. The θ = 0 data are the time-resolved PL when no microwaves are applied and the spin remains in $m_s$ = 0, while the θ = π data are taken after a microwave π-pulse prepares the spin in the $m_s$ = -1 state. The small gaps originate from a preprocessing step where small pulses imperfectly extinguished by our electro optic modulator are removed. (c) Normalized single divacancy photoluminescence

emitted at the beginning off-resonant CW laser pulses of different power. The upper trace in each subpanel is the PL observed when the spin is prepared in $m_s$ = 0, while the respective lower traces are observed when a microwave π-pulse rotates the spin into $m_s$ = -1. Solid lines are the corresponding global fits of the rate-equation model.